\documentclass[prb,twocolumn,aps,showpacs,fixfloats]{revtex4}
\usepackage{graphicx}
\usepackage{bm}
\usepackage{amsmath,amssymb}
\usepackage{subfigure}
\usepackage{float}
\usepackage{latexsym}
\usepackage{epstopdf}
\usepackage{color}
\usepackage{enumerate}

\begin{document}
\newcommand{\s}{\scriptscriptstyle}
\newcommand{\uu}{\uparrow \uparrow}
\newcommand{\ud}{\uparrow \downarrow}
\newcommand{\du}{\downarrow \uparrow}
\newcommand{\dd}{\downarrow \downarrow}
\newcommand{\ket}[1] { \left|{#1}\right> }
\newcommand{\bra}[1] { \left<{#1}\right| }
\newcommand{\bracket}[2] {\left< \left. {#1} \right| {#2} \right>}
\newcommand{\vc}[1] {\ensuremath {\bm {#1}}}
\newcommand{\tr}{\text{Tr}}

\title{Spin dynamics and spin-dependent recombination in a polaron pair under a strong ac drive}

\author{R. K. Malla  and M. E. Raikh} \affiliation{Department of Physics and
Astronomy, University of Utah, Salt Lake City, UT 84112}

\begin{abstract}
We study theoretically the recombination within a pair of two polarons
in magnetic field subject to a strong linearly polarized ac drive. Strong
drive implies that the Zeeman frequencies of the pair-partners are much
smaller than the Rabi frequency, so that the rotating wave approximation does not apply. What makes the recombination dynamics nontrivial, is that the partners recombine only when they form a singlet, $S$.
By admixing singlet to triplets, the drive induces the triplet recombination as well. We calculate the effective decay rate of all four spin modes. Our main finding is that, under the strong drive, the major contribution to the decay of the modes comes from short time intervals when the driving field passes through zero.  When the recombination time in the absence of drive is short, fast recombination from $S$ leads to anomalously slow recombination
from the other spin states of the pair. We show that, with strong drive, this recombination becomes even slower. The corresponding decay rate falls off as a power law with the
amplitude of the drive.
\end{abstract}

\pacs{73.50.-h, 75.47.-m}
\maketitle

\section{Introduction}
The dynamics of a spin in an external magnetic field, ${\bm B}$, is governed by the
equation ${\dot {\bm S}}={\bm B}\times {\bm S}$. In the  magnetic resonance setup
one has ${\bm B}={\bm z}_0B+{\bm x}_0B_1\sin\omega t$, where $B$ is the constant
field, while $B_1$ and $\omega$ are the amplitude and the frequency of drive.
It is easy to see that, in the general case, the dynamics is governed by two dimensionless parameters, $B_1/\omega$ and $B/\omega$.

\begin{figure}
\includegraphics[scale=0.7]{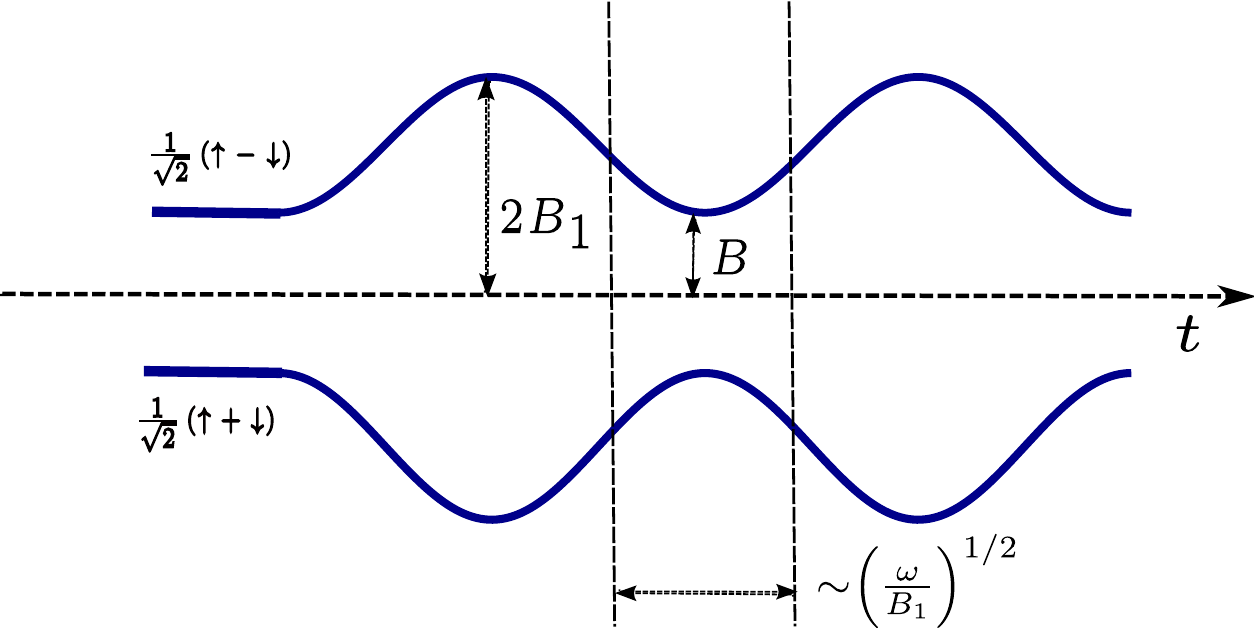}
\caption{(Color online) Schematic illustration of the adiabatic energy levels
of a strongly driven single spin. The dc field, $B$, is along the $z$-axis, while the drive field with amplitude $B_1\gg B$ and frequency $\omega$ is along $x$. Non-adiabatic  spin dynamics takes place in the narrow, $\sim \left(\omega/B_1\right)^{1/2}$,  time
 domains where the drive passes through zero. Between these zeros the precession frequency is determined by
the instant value of the driving field.}
\label{p1}
\end{figure}

For a resonant drive, $B \approx \omega$, the dynamics represents conventional Rabi oscillations\cite{Rabi0}, with frequency $B_1$. They take place for a weak
drive $B_1\ll \omega$. Upon increasing $B_1$, the resonant condition gets
modified to $\omega\approx B+B_1^2/\omega$ due to the Bloch-Siegert shift\cite{Rabi0'}. The study of the spin dynamics under the conditions  when the two dimensionless
parameters take arbitrary values was pioneered in the seminal papers
Refs. \onlinecite{Rabi1}, \onlinecite{Rabi2}. In particular, the analytical
results for the Floquet exponent, which is the prime characteristics of the dynamics, was obtained in Ref. \onlinecite{Rabi2} in the limit
$\frac{B_1}{\omega}\gg 1$, while  $\frac{B}{\omega}\sim 1$.
Recently, this regime of a very strong drive became relevant in actively
developing field of superconducting qubits, namely in the Landau-Zener-Stueckelberg interferometry, see e.g.
Refs. \onlinecite{LZ1},  \onlinecite{LZ2}, and the review Ref. \onlinecite{LZ3}.
The physical picture of the dynamics under the strong drive is that the spin rapidly precesses
around the instant value of the driving field. As a result, the spin projections oscillate not as $\cos (Bt)$, like in a constant field,
but as $\cos\left(\int B_x(t)dt\right)$, i.e. as
$\cos\left(\frac{B_1}{\omega}\cos \omega t  \right)$. Thus, the number of oscillations during the period, $2\pi/\omega$, is equal to
$\frac{1}{2\pi}\left(\frac{4B_1}{\omega}\right)$, i.e.  it is large. This justifies the above physical picture.
Obviously, the rapid  precession around the drive field is interrupted within narrow time domains around $\omega t \approx \pi n$, as illustrated in Fig. \ref{p1}. At these
 moments the drive is small. A spin passes these domains by
 undergoing the Landau-Zener transitions.\cite{LZ3,LZ4}

 In different realizations of superconducting qubits on which Landau-Zener-Stueckelberg interferometry experiments were carried out, the role of the field, $B$, responsible for the spacing between upper and lower energy states, can be played by different parameters.\cite{LZ3} The role of the driving field is played by the magnetic flux, which modulates the Josephson energy. Most importantly, the ratio $B_1/\omega$ can be varied in a wide range, see e.g. Refs. 9-11.

In the present paper we focus on a completely different system in which
the spin dynamics under a strong drive can be detected by electrical means. This system is an ensemble of polaron pairs in organic materials.
Dynamics of a single polaron in magnetic field and ac drive is a conventional two-level system dynamics which is detected by electron paramagnetic resonance. However, electrical detection relies on recombination of {\em two} polarons.\cite{Kaplan} One approach to this detection,
justified theoretically in Ref. \onlinecite{Recombination0} and realized
experimentally in a number of papers,\cite{Recombination1,Recombination2,Recombination21,Recombination3,Recombination4,Recombination5,Recombination6,Recombination7,Recombination8}
is pulsed electrically detected magnetic resonance. In this technique, the net charge passed through a sample is measured as a function of duration of the ac drive pulse. The reason why this charge reflects the spin dynamics
during the pulse is the spin-dependent recombination. More specifically,
the initial and the final states of the pair can be either $|\downarrow\downarrow\rangle$ or $|\uparrow\uparrow\rangle$. Any admixture of a singlet forces the pair to quickly recombine. Thus it is the probability for a pair to have the ``right" initial and final states that
determines the change of the bulk conductivity. This probability is sensitive to the dynamics of the pair partners.
As a result, the Fourier
transform of
the transmitted charge
with respect to the pulse duration exhibits
the peaks corresponding to the Rabi oscillations frequencies.\cite{Recombination1,Recombination2,Recombination21,Recombination3,Recombination4,Recombination5,Recombination6,Recombination7,Recombination8}
The other phenomenon which fully relies on the spin-dependent recombination
(or, alternatively,  bipolaron formation) is organic magnetoresistance.\cite{Dediu,Valve,Markus0,Markus1,Markus2,Prigodin,Gillin,Bobbert1,BobbertStochastic,
XuWu,Valy0,Blum,Valy1,Flatte1,we,we1,fringe,Harmon} In bipolar devices,
recombination of electron and hole polarons injected from the electrodes is responsible for the passage of current. If the spin state of a pair, assembled at neighboring sites, does not have a singlet admixture, the pair will never recombine (spin blockade). Sensitivity of the current through the device to external magnetic field is caused by  redistribution of the number of the blocking pairs. As it was demonstrated in Refs. \onlinecite{Boehme0}-\onlinecite{Boehme1}, ac drive strongly affects the
current when its frequency is near the resonance. The underlying reason for this
is lifting the spin blockade.

So far, all the experiments on spin manipulation of the pairs by ac drive in organic materials were carried out near the resonant condition $\omega \approx B$. This is because realization of the strong-drive regime was precluded by the random hyperfine fields on the sites where the pair-partners resided. A typical magnitude of these fields is $\sim 1$mT.
In the experiments on pulsed magnetic resonance\cite{Recombination1,Recombination2,Recombination21,Recombination3,Recombination4,Recombination5,Recombination6,Recombination7,Recombination8}
the drive frequency was in the GHz domain, and correspondingly, the field $B$ was of the order of $100$mT, much bigger than $B_1$. However, in experiments\cite{Boehme1,Baker} on magnetoresistance under the ac drive with frequency $\sim 100$MHz, the field $B\approx 3$mT was several times bigger than the hyperfine field and only $3$ times bigger than $B_1$.  For this setup achieving the strong-drive regime seems feasible.

The spin dynamics of a strongly driven pair studied in the present paper is much richer than the dynamics of a single spin. The reason is that,
with recombination allowed only from the {\em entangled}  singlet state,
the dynamics of the pair partners becomes coupled {\em via recombination}.
A dramatic consequence of this coupling is emergence of long-leaving modes\cite{we3,Boehme1} with
decay time much longer than the lifetime of a singlet. These modes are similar
to subradiant modes in the Dicke effect\cite{Dicke}.

%
%

The paper is organized as follows. In Sect. II we introduce the system of equations of motion
for a driven spin pair. In Sect. III we present the solutions of this system in the limit of
very long recombination time. This solutions are derived in Sect. IV and analyzed in Sect. V. In Sect. VI we consider finite recombination time and calculate effective lifetimes of all the modes of spin dynamics of the driven pair. In Sect. VII we summarize our main findings qualitatively.
Concluding remarks are presented in Sect. VIII.

\begin{figure}
\includegraphics[scale=0.35]{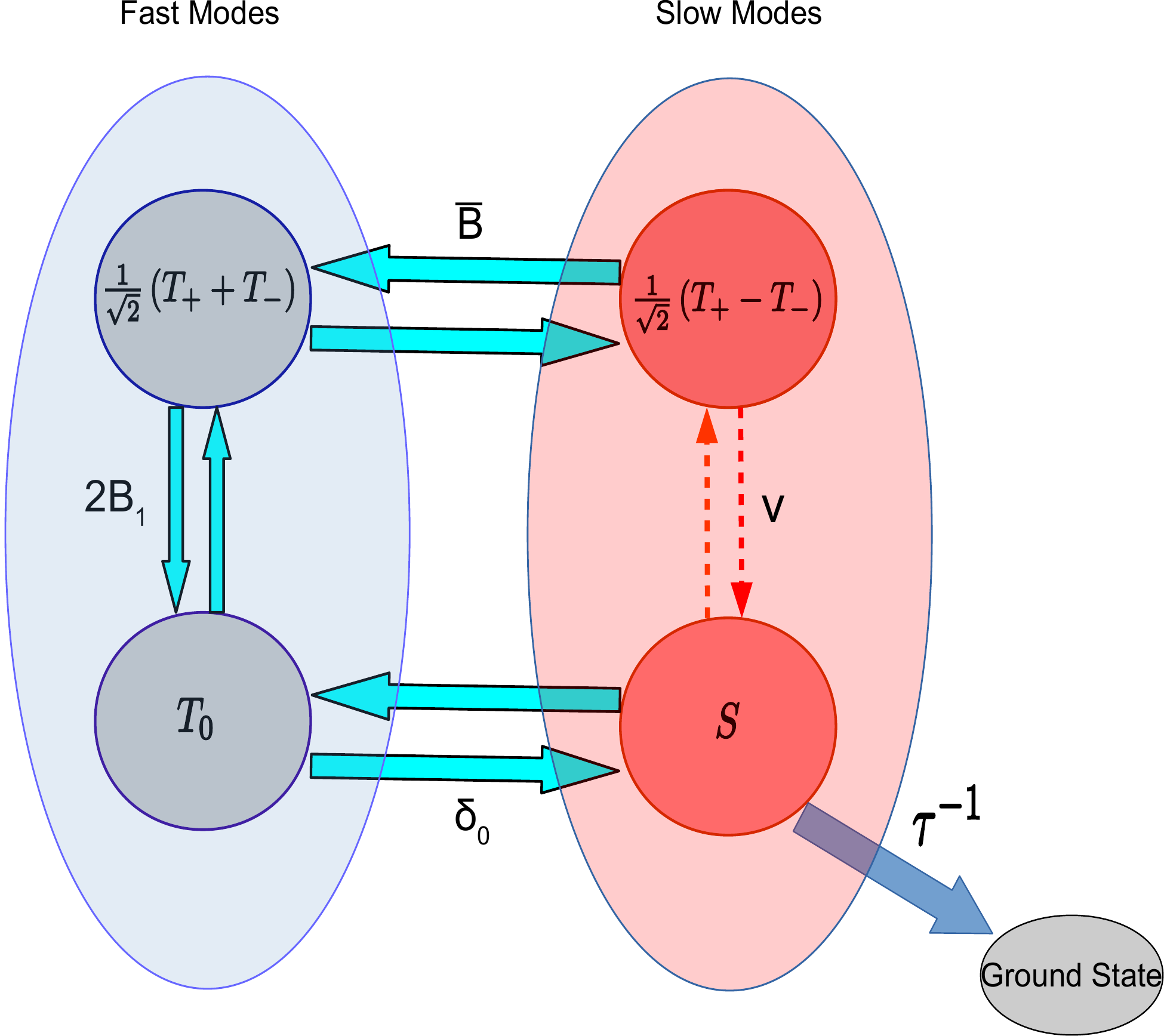}
\caption{(Color online) Formation of slow and fast modes in the dynamics of a spin pair
under a strong drive is illustrated schematically. In the absence of drive,
the eigenmodes correspond to  $\frac{1}{\sqrt 2}\left(T_{+}+T_{-}\right)$ coupled to
$\frac{1}{\sqrt 2}\left(T_{+}-T_{-}\right)$ via the average field,
$\overline{B}$, and to  $S$ coupled to $T_{0}$ via the difference of the hyperfine fields, $\delta_0$. Strong drive introduces the coupling between $\frac{1}{\sqrt 2}\left(T_{+}+T_{-}\right)$ and $T_{0}$ giving rise to the 
fast modes Eq. (\ref{fastmatrix}). The remaining two states $\frac{1}{\sqrt 2}\left(T_{+}-T_{-}\right)$ and $S$ constitute slow modes. The coupling
between these states, quantified by parameter $\nu$ Eq. (\ref{nu}), remains weak as in the absence of the drive. With recombination from $S$,
fast modes decay weakly only during the moments when the drive passes through zero. The recombination of the slow modes depends on the relation between
$\delta_0$ and $\tau^{-1}$.  }
\label{p2}
\end{figure}

\section{ac-driven spin-pair with recombination}

The Hamiltonian of the pair in a linear polarized driving field with amplitude, $B_1$, reads
\begin{align}
{\hat H} = B_aS_{a}^{z}+B_bS_{b}^{z}+2B_1\left(S_{a}^{x}+S_{b}^{x}\right)\sin \omega t,
\end{align}
where $\omega$ is the driving frequency, while $B_a$ and $B_b$ are the net fields (in the frequency units) acting on the partners $a$ and $b$, respectively.


If the hyperfine fields acting on both components are the same, then
the dynamics of a pair is trivial. The
initial state, $S$, decays with recombination time, $\tau$, while other initial states $T_{+}$, $T_{-}$, and $T_0$ do not decay at all. Finite
\begin{equation}
\label{delta0}
\delta_0=\frac{B_a-B_b}{2}
\end{equation}
leads to the mixing of the amplitudes of $S$ and $T_0$, so that the amplitudes of the $4$
states are  related by the system
\begin{align}
i\frac{\partial T_{+}}{\partial t} &= \overline{B} T_{+}+\sqrt{2}B_1T_0\sin \omega t,\label{drivensystem1}  \\
i\frac{\partial T_{-}}{\partial t}&=-\overline{B} T_{-}+\sqrt{2}B_1T_0\sin \omega t,\label{drivensystem2}\\
i\left( \frac{\partial S}{\partial t}+\frac{S}{\tau}\right) &=\delta_0 T_{0},\label{drivensystem3}\\
i\frac{\partial T_{0}}{\partial t}&=\delta_0 S+\sqrt{2}B_1\left(T_{+}+T_{-}\right)\sin \omega t, \label{drivensystem4}
\end{align}
where
\begin{equation}
\label{Baverage}
\overline{B}=\frac{B_a+B_b}{2}
\end{equation}
is the average  $z$-component of the net fields.
The above system of $4$ equations for the amplitudes
is equivalent to the system $16$
equations for the elements of the density matrix
and constitute a starting point of numerous studies of magnetic
resonance with pair-recombination. Unlike the present paper, the
weak drive limit, $B_1 \ll {\overline B}$, is implied in all the
earlier studies.

\section{The content of eigenmodes}
 Conventionally, the system Eqs.~(\ref{drivensystem1})-(\ref{drivensystem4}) is analyzed within the rotating wave approximation which applies when the drive is weak compared to ${\overline B}$. In the rotating wave approximation the eigenmodes of the system the
system represent the product of the sinusoidal functions with frequencies
$\Big[\left(\overline{B}+\delta_0-\omega\right)^2+B_1^2\Big]^{1/2}$ and $\Big[\left(\overline{B}-\delta_0-\omega\right)^2+B_1^2\Big]^{1/2}$. As the drive amplitude
increases, the sum of the frequencies approaches $2B_1$, while their difference becomes much
smaller than $B_1$. Then the eigenmodes can be classified into ``fast" and ``slow". We consider the opposite limit $B_1 \gg {\overline B}$.  Classification of the eigenmodes into
fast and slow still applies in this limit, see Fig \ref{p2}. Below we present our result for the form of the eigenmodes, while the derivation is outlined later on.

The solution of the system Eq. (\ref{drivensystem1})-Eq. (\ref{drivensystem4}) for the two
fast eigenmodes has the form
{\small \begin{eqnarray}
\label{fastmatrix}
\quad\begin{pmatrix}
  T_+\\ \\T_- \\ \\T_0\\ \\S \end{pmatrix}=\frac{\exp\Big[\pm i\varphi_{\s f}(t)\Big]}{4}\hspace{-1mm}\begin{pmatrix} \\1-\frac{1}{2B_1 \sin \omega t}\left(\pm i\frac{\omega \cos \omega t}{\sin \omega t}-\overline{B}\right)\\ \\1-\frac{1}{2B_1 \sin \omega t}\left(\pm i\frac{\omega \cos \omega t}{\sin \omega t}+\overline{B}\right)  \\ \\ \mp\sqrt{2}\left(1\mp i\frac{\omega \cos \omega t}{2B_1 \sin^2 \omega t}\right) \\ \\ \pm \sqrt{2}\left(\frac{\delta_0}{2B_1 \sin \omega t}\right) \end{pmatrix}\hspace{0.5mm}
\end{eqnarray}}
where the``fast" phase $\varphi_{\s f}$ is defined as
\begin{equation}
\label{fastphase}
\varphi_{\s f}(t)=\frac{2B_1}{\omega}\cos \omega t +\mu \ln\left(\frac{1-\cos \omega t}{1+\cos \omega t}\right),
\end{equation}
and parameter $\mu$ is defined as
\begin{equation}
\label{mu}
\mu = \frac{\overline{B}^2+\delta_0^2}{4B_1 \omega}.
\end{equation}
The expression Eq. (\ref{fastmatrix}) applies in the ``semiclassical" limit when $\varphi_{\s f}$ is bigger than one, i.e. when
\begin{equation}
\label{condition}
\frac{B_1}{\omega}|\omega t-\pi n|^2 \gg 1.
\end{equation}
The above condition defines a narrow, $\sim\left(\omega/B_1\right)^{1/2}$,  domain around $\omega t=\pi n$, where the semiclassical description is not applicable. Note that, within the allowed domain, the phase $\varphi_{\s f}$ accumulated within a period of drive is $2B_1/\omega$, i.e. it is big. The condition Eq. (\ref{condition}) also ensures that the terms containing  $\sin \omega t$
in the denominators in Eq. (\ref{fastmatrix}) do not exceed the main terms. In particular, the
maximum value of the $S$-component is $\sim \delta_0/\left(\omega B_1\right)^{1/2} \ll 1$.

The spinors describing two slow modes have the form

\begin{eqnarray}
\label{slowmatrix}
\begin{pmatrix}
 T_+\\ \\T_-\\ \\T_0\\ \\S \end{pmatrix}=\frac{\exp\Big[\pm i\varphi_{s}(t)\Big]}{4}\begin{pmatrix} 1\mp\frac{\delta_0}{2B_1\sin \omega t} \\ \\-1\mp\frac{\delta_0}{2B_1\sin \omega t} \\ \\\mp\sqrt{2}\frac{\overline{B}}{2B_1\sin \omega t} \\ \\\pm\sqrt{2} \end{pmatrix},
\end{eqnarray}
where $\varphi_s$ is the slow phase given by
\begin{equation}
\label{slowphase}
\varphi_{s}(t)=\nu \ln\left(\frac{1-\cos \omega t}{1+\cos \omega t}\right),
\end{equation}
and parameter $\nu$ is defined as
\begin{equation}
\nu = \frac{\delta_0\overline{B}}{2B_1\omega}.\label{nu}
\end{equation}
The condition Eq. (\ref{condition}) guarantees that the terms containing $\sin \omega t$ in the
denominator constitute small corrections to the main terms. The corrections are of the order of $\delta_0/\left(B_1\omega\right)^{1/2}$ for $T_{+},~T_{-}$ and of the order of ${\overline B}/\left(B_1\omega\right)^{1/2}$  for $T_0$.  We also note that, with $\nu$ being a small parameter, and with divergence in the argument of
logarithm being cut off, the phase $\varphi_s$ does not exceed $1$.

\section{Derivation}
We start the derivation by reducing the system of four first-order differential equations  Eq.~(\ref{drivensystem1})-(\ref{drivensystem4}) to two second-order differential equations.
Upon adding and subtracting Eqs. (\ref{drivensystem1}) and  (\ref{drivensystem2}) we get
\begin{align}
i\frac{\partial \left(T_{+} + T_{-}\right)}{\partial t} &= \overline{B}\left( T_{+}-T_{-}\right)+2\sqrt{2}B_1T_0\sin \omega t,\label{newdrivensystem1} \\
i\frac{\partial \left(T_{+} - T_{-}\right)}{\partial t} &= \overline{B}\left( T_{+}+T_{-}\right).\label{newdrivensystem2}
\end{align}
As a next step, we substitute $T_0$ from Eq. (\ref{drivensystem3}) and $\left(T_{+}+T_{-}\right)$ from Eq. (\ref{newdrivensystem2}) into Eq. (\ref{drivensystem4}). This yields
\begin{align}
\frac{\partial^2S}{\partial t^2}+\frac{1}{\tau}\frac{\partial S}{\partial t}+\delta_0^2 S &= -i\sqrt{2}~B_1\frac{\delta_0}{\overline{B}}\sin \omega t\frac{\partial \big(T_{+}-T_{-}\big)}{\partial t}. \label{differential1}
\end{align}
Finally, we substitute $T_0$ from Eq. (\ref{drivensystem3})  and  $\left(T_{+}+T_{-}\right)$ from Eq. (\ref{newdrivensystem2}) into Eq. (\ref{newdrivensystem1}) and obtain

\begin{multline}
\frac{\partial^2\big(T_{+}-T_{-}\big)}{\partial t^2} + \overline{B}^2 \big(T_{+}-T_{-}\big) \\ = -2i\sqrt{2}B_1\frac{\overline{B}}{\delta_0}\sin \omega t\left(\frac{\partial S}{\partial t} +\frac{S}{\tau}\right).
\label{differential2}
\end{multline}
From the solution of the system Eqs. (\ref{differential1}),  (\ref{differential2}) the amplitude $T_0$ can be found
using Eq. (\ref{drivensystem3}), while $T_{+}$ and $T_{-}$ can be found from Eqs.  (\ref {newdrivensystem1}) and (\ref {newdrivensystem2}).

Solution of the system
corresponding to
slow modes emerges upon neglecting second derivatives and setting $\tau \rightarrow \infty$.
Then it takes the form
\begin{align}
i\nu S &= \frac{1}{\sqrt{2}}\sin \omega t\hspace{1mm}\frac{\partial \big(T_{+}-T_{-}\big)}{\omega \partial t},\label{slowdifferential1}\\
i\nu\big(T_{+}-T_{-}\big) &= \sqrt{2}\sin \omega t\hspace{1mm}\frac{\partial S}{\omega \partial t},\label{slowdifferential2}
\end{align}
where $\nu$ is defined by Eq. (\ref{nu}). Note now, that $\sin \omega t \frac{\partial}{\partial t}$ can be rewritten as $\partial/\partial \ln\left(\frac{1-\cos \omega t}{1+\cos \omega t}\right)$. This immediately suggests that $S(t) \propto \exp{(\pm i\varphi_{s}(t))}$ in accordance with Eq. (\ref{slowphase}). Neglecting second derivatives is justified by smallness of the parameter $\nu$ and the condition Eq. (\ref{condition}). Indeed, using Eq. (\ref{slowphase}), we find
\begin{align}
\label{secondderivative}
\frac{1}{\delta_{0}^{2}S}\left(\frac{\partial^2S}{\partial t^2}\right)=\frac{1}{\sin^2 \omega t}\left[\left(\frac{\overline{B}}{2B_1}\right)^2 + i \left(\frac{\overline{B}}{\delta_0}\right)\left(\frac{\omega}{2B_1}\right)\cos \omega t\right].
\end{align}
The left-hand side is the ratio of the neglected term to the term we kept in Eq. (\ref{slowdifferential1}). We see that, since  $\left(\sin \omega t\right)^{-2}$ cannot exceed
$B_1/\omega$
according to the condition Eq. (\ref{condition}), both terms in the right-hand side are small.
For slow modes the $\big(T_{+}-T_{-}\big)$ and $S$ components of spinors are related as
$\big(T_{+}-T_{-}\big)=\pm \sqrt{2}S$, as reflected in Eq. (\ref{slowmatrix}).

Turning to the fast modes, instead of the variables $S(t)$, $\big(T_{+}(t)-T_{t}(t)\big)$, we introduce new variables, ${\tilde S}(t)$ and ${\tilde T}(t)$, in the following way
\begin{eqnarray}
\label{SlowSemiclassics}
\begin{pmatrix}
  S \\  \\ T_{+}-T_{-} \end{pmatrix}=\frac{\exp\left(i\frac{2B_1}{\omega}\cos\omega t\right)}{\sin \omega t \left(\delta_0^2+\overline{B}^2\right)^{1/2}}\begin{pmatrix} \delta_0 {\tilde S}(t) \\  \\ \overline{B}{\tilde T}(t)
\end{pmatrix}.
\end{eqnarray}
When substituting these new variables in Eqs. (\ref{differential1}), (\ref{differential2}) we assume  that $\tilde{S}$ and $\tilde{T}$ are slow functions and neglect their second derivatives. We also take into account that, by virtue of the condition Eq. (\ref{condition}), the  derivative of $1/\sin \omega t$ is much smaller than the derivative of the exponent. Then the system takes the form
\begin{equation}
\label{systemnewform}
2\dot{\tilde{S}}-\dot{\tilde{T}}=-i\frac{\delta_0^2}{2B_1\sin \omega t}\tilde{S}+\left(2iB_1\sin \omega t+\omega \cot \omega t\right)\left(\tilde{S}-\tilde{T}\right),
\end{equation}
\begin{equation}
\label{systemnewform1}
2\dot{\tilde{T}}-\dot{\tilde{S}}=-i\frac{\overline{B}^2}{2B_1\sin \omega t}\tilde{T}+\left(2i B_1\sin \omega t+\omega \cot \omega t\right)\left(\tilde{T}-\tilde{S}\right).\\
\end{equation}
Upon adding and subtracting Eqs. (\ref{systemnewform}), (\ref{systemnewform1}), we find
\begin{equation}
\label{systemnewformadd}
\dot{\tilde{T}}+\dot{\tilde{S}}=-i\frac{\delta_0^2\tilde{S}+\overline{B}^2\tilde{T}}{2B_1\sin \omega t },
\end{equation}
\begin{multline}
\label{systemnewformsub}
3\left(\dot{\tilde{T}}-\dot{\tilde{S}}\right)=-i\frac{\delta_0^2\tilde{S}-\overline{B}^2\tilde{T}}{2B_1\sin \omega t } \\ + 2\left(2i B_1\sin \omega t+\omega \cot \omega t\right)\left(\tilde{T}-\tilde{S}\right).
\end{multline}
It follows from Eq. (\ref{systemnewformsub}) that the solution for $\tilde{T}-\tilde{S}$ is given by
\begin{equation}
\label{conditionTS}
\tilde{T}-\tilde{S}\approx \frac{\delta_0^2\tilde{S}-\overline{B}^2\tilde{T}}{8B_1^2\sin^2 \omega t }.
\end{equation}
This solution is the result of neglecting the left-hand side.
The correction from finite left-hand side is small by virtue of condition Eq. (\ref{condition}). With difference $\tilde{T}-\tilde{S}$ being of the order of $1/B_1^2$, we can set $\tilde{T} = \tilde{S}$ in Eq. (\ref{systemnewformadd}) and obtain

\begin{equation}
\tilde{S}=\tilde{T}=\frac{1}{2}\exp\Bigg[-i\frac{\delta_0^2+\overline{B}^2}{4B_1\omega}\ln\left(\frac{1-\cos \omega t}{1+ \cos \omega t}\right)\Bigg].
\end{equation}
We see the $\tilde{S}$ and $\tilde{T}$ are indeed slow functions, as was assumed above.
Returning back to $S$ and $T$, we recover the results Eqs. (\ref{fastmatrix}), (\ref{fastphase}).



\section{Applications}
Below we address the question of the time evolution of the pair created at some moment, $t_0$, in
one of the states $T_{+}$, $T_{-}$, $T_0$, or $S$. We start from $T_{+}$, and assume that at $t=t_0$ the amplitude of $T_{+}$ is $1$, while the amplitudes of other states are zero. These conditions are
satisfied by a certain linear combination of eigenmodes Eqs. (\ref{fastmatrix}, (\ref{slowmatrix}).
It is easy to see that, in the limit $B_1\gg {\overline B}, \delta_0$, the coefficients are
$\exp{\left(\pm \varphi_{\s f}(t_0)\right)}$ and $\exp{\left(\pm \varphi_s(t_0)\right)}$.
This leads to the following time evolution of $T_{+}$
\begin{align}
\label{t1evolution}
T_{+}(t)=\frac{1}{2}\Big[ \cos \left(\varphi_{\s f}(t)-\varphi_{\s f}(t_0)\right)+ \cos \left(\varphi_s(t)-\varphi_s(t_0)\right)\Big].
\end{align}
Neglecting the logarithmic corrections, we can set $\varphi_{\s f}(t)=\frac{2B_1}{\omega}\cos \omega t$ and $\varphi_s(t)=0$. This leads to a simple expression for the probability to find the pair
in the state $T_{+}$  at time $t$
\begin{equation}
\label{t1probability}
|T_{+}(t)|^2=\frac{1}{8}\big[3+\cos 2\theta(t) + 4\cos \theta(t)\big],
\end{equation}
where the $\theta(t)$ is defined as
\begin{equation}
\theta(t)=\frac{2B_1}{\omega}(\cos \omega t - \cos \omega t_0).
\end{equation}
The corresponding probability to find the pair in $T_{-}$ reads
\begin{equation}
\label{t2probability}
|T_{-}(t)|^2=\frac{1}{8}\big[3+\cos 2\theta(t) - 4\cos \theta(t)\big].
\end{equation}
The normalization is ensured by contribution of $T_0$ for which the dynamics contains only
a double frequency, $2\varphi_{\s f}(t)$, namely
\begin{equation}
\label{t3probability}
|T_{0}(t)|^2=\frac{1}{4}\big[1-\cos 2\theta(t) \big].
\end{equation}
In calculating the dynamics we neglected the slow oscillations with frequency $\varphi_s(t)$.
These slow oscillations govern the dynamics of $S$. The probability, $|S(t)|^2$ has the form
\begin{multline}
\label{sprobability}
|S(t)|^2=\frac{1}{4}\Big\{1-\cos \Big[2\big( \varphi_s(t)-\varphi_s(t_0)\big)\Big]\Big\}.
\end{multline}
\begin{figure}
\includegraphics[scale=0.75]{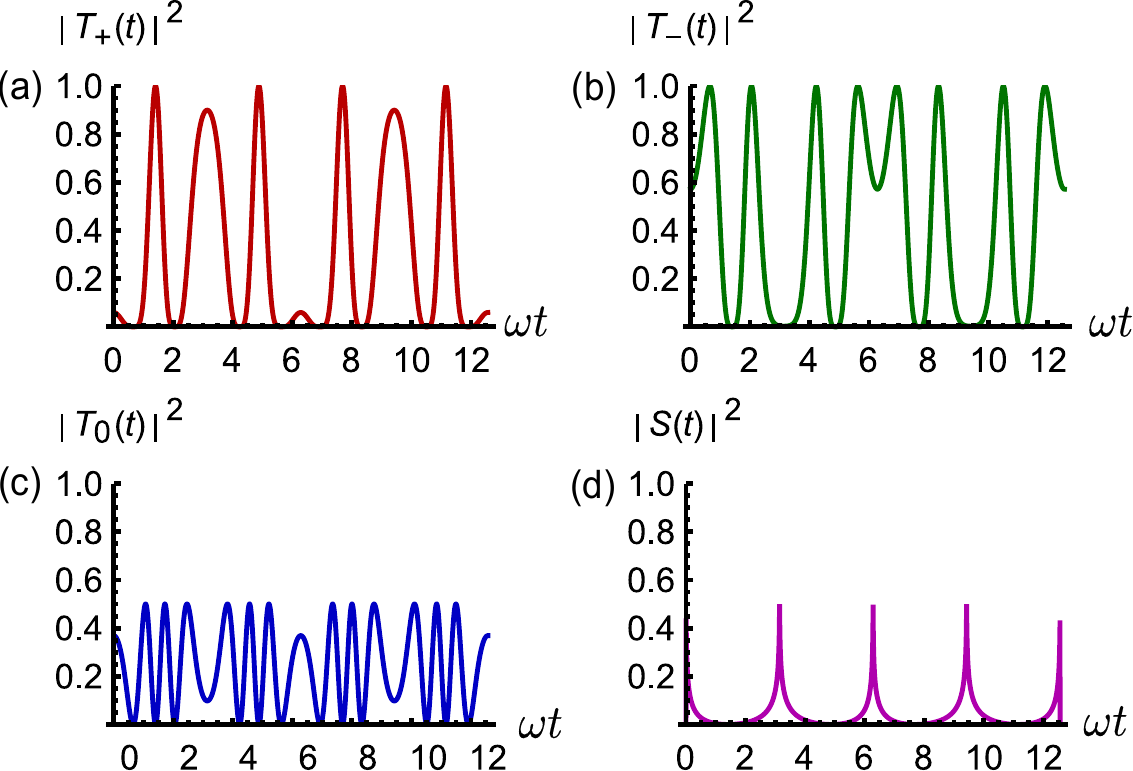}
\caption{(Color online) At time moment $t=t_0=\pi/\sqrt{5}\omega$ the system is in the state $T_{+}$. The evolution of the occupations of $T_{+}$ (a), $T_{-}$ (b), $T_{0}$ (c),
and $S$ (d) is plotted from Eqs. (\ref{t1probability})-(\ref{sprobability})
for the drive amplitude $2B_1/\omega = 5$. Due to the small value of parameter $\nu=0.1$, the magnitude of the $S$ component is small and it changes only in the vicinity of $\omega t=\pi n$.  The dynamics shown in the figure should be contrasted to the conventional Rabi oscillations,
when $B_1$ is much smaller than $\omega\approx \overline{B}$. }
\end{figure}
Smallness of parameter $\nu$ in the expression for $\varphi_s(t)$ allows to simplify this expression to
\begin{multline}
\label{sprobabilitysimple}
|S(t)|^2=\frac{\nu^2}{2}\Bigg[\ln\left(\frac{1-\cos \omega t}{1-\cos \omega t_0}\right)-\ln\left(\frac{1+\cos \omega t}{1+\cos \omega t_0}\right)\Bigg]^2.
\end{multline}
With regard to  observables, the dynamics of, say, $T_{+}(t)$ manifests itself in the
Fourier spectrum $\bm{F}(s)=\int dt|T_{+}(t)|^2\cos st$. The spectrum is the set of
$\delta$-peaks at $s=n\omega$. The magnitudes, $\bm{F}_n$, of these peaks calculated from
Eq. (\ref{t1probability}) are given by
\begin{multline}
\label{t1fourier}
\bm{F}_{n}= \Bigg[\cos\left(\frac{2B_1}{\omega}\cos \omega t_0 -\frac{n\pi}{2}\right)J_n\Big(\frac{2B_1}{\omega}\Big)\\+\frac{1}{4}\cos\left(\frac{4B_1}{\omega}\cos \omega t_0 -\frac{n\pi}{2}\right)J_n\Big(\frac{4B_1}{\omega}\Big)\Bigg],
\end{multline}
\begin{figure}
\includegraphics[scale=0.60]{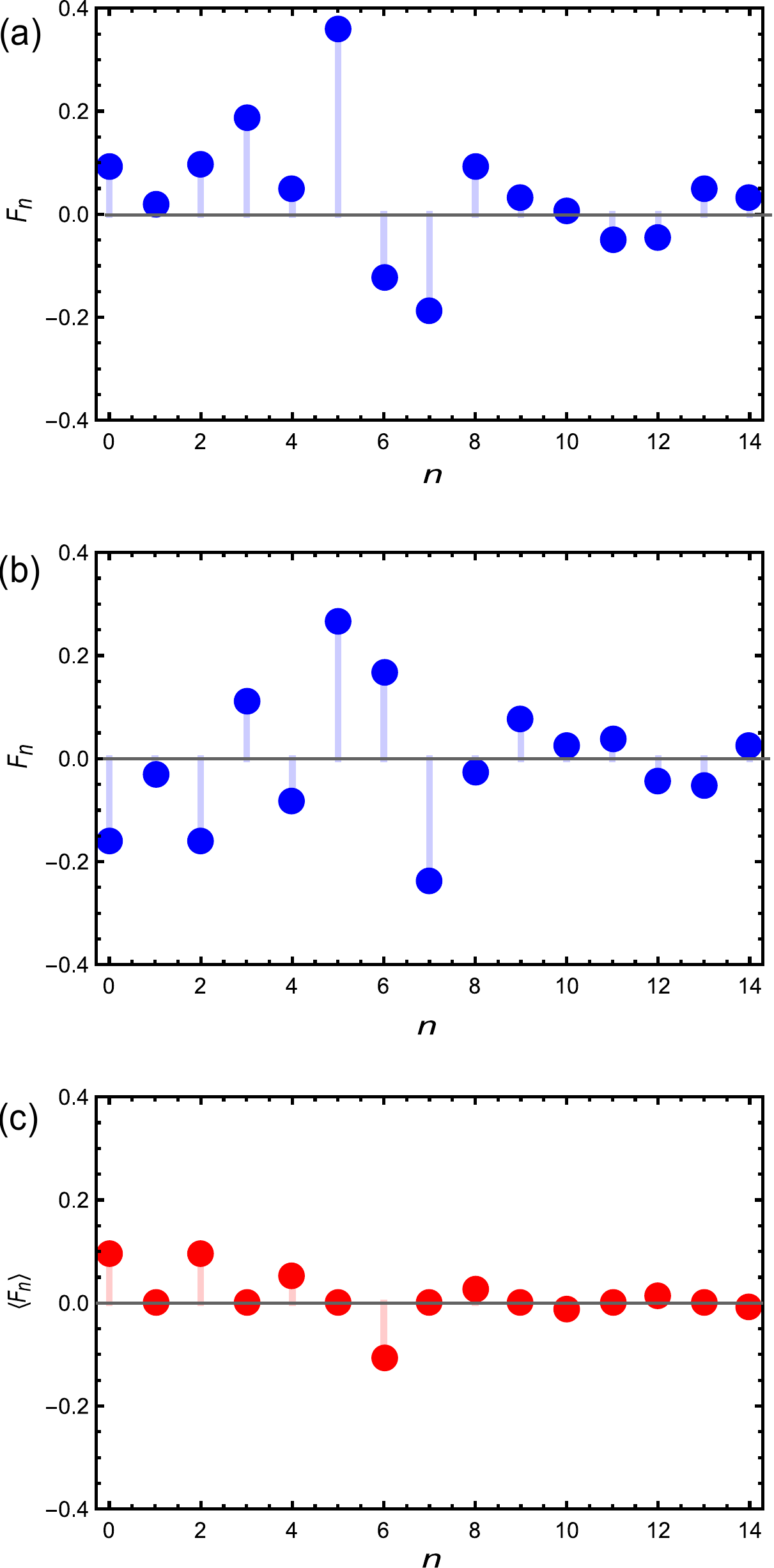}
\caption{(Color online) At time moment $t=t_0$ the system is in the state $T_{+}$. The spectrum of the $T_{+}$ occupation is plotted from Eq. (\ref{t1fourier}) for the values: (a) $\omega t_0= \pi/\sqrt{5}$,
and (b) $\omega t_0=
\pi/\sqrt{2}$. Remarkably, as shown in the panel (c) plotted from Eq. (\ref{averagefourier}), the features in the spectrum
survive after averaging over $t_0$. The plots are for the ratio
$2B_1/\omega =7$.  }
\label{p4}
\end{figure}
where $J_n(x)$ is the Bessel function. It follows from Eq. (\ref{t1fourier})
that the spectrum depends very strongly on the phase of the drive at the moment of the pair formation. Since $J_n(x)$
drops sharply when $n$ exceeds $x$, we conclude that the number of peaks in the spectrum with appreciable magnitudes is, essentially, $4B_1/\omega$, i.e. with increasing the drive amplitude the
spectrum becomes progressively rich. The shapes of the spectrum for different $\omega t_0$  are illustrated in Fig. \ref{p4}.

More relevant to experiment\cite{Recombination1,Recombination2,Recombination21,Recombination3,Recombination4,Recombination5,Recombination6,Recombination7,Recombination8}  can be the Fourier transform averaged over the moments, $t_0$, of the pair formation. The averaging of Eq. (\ref{t1fourier}) is straightforward and yields
\begin{multline}
\label{averagefourier}
\langle \bm{F}_n \rangle_{t_0}=\cos \frac{n\pi}{2}\Bigg[J_0\left(\frac{2B_1}{\omega}\right)J_n\left(\frac{2B_1}{\omega}\right) \\ +\frac{1}{4}J_0\left(\frac{4B_1}{\omega}\right)J_n\left(\frac{4B_1}{\omega}\right)\Bigg].
\end{multline}
As illustrated in Fig. \ref{p4}, the averaged spectrum retains a lively
structure.

Unlike Eq. (\ref{t1fourier}), the spectrum of $S(t)$  depends on the phase of the drive only weakly. The expression for the magnitudes, ${\bm G}_n$, of the peaks takes a simple form for $n \gg 1$, namely,
\begin{equation}
\label{sfourier}
{\bm G}_n=8\nu^2\frac{\ln n}{n}.
\end{equation}

According to Eq. (\ref{sfourier}) the magnitudes of the peaks fall off with $n$ quite slowly.
However, this behavior is terminated at large $n$. This is because, for large $n$, the Fourier
components are determined by a narrow time domain $\omega t \sim 1/n$. On the other hand, according
to Eq. (\ref{condition}), the value $\omega t$ cannot be smaller than $(\omega/B_1)^{1/2}$.
Thus, the maximal $n$ can be estimated as $n_{\text{max}}\sim (B_1/\omega)^{1/2}$.

Assume now, that at $t=t_0$ the pair is created in the state $S$.
Again, in the limit $B_1\gg {\overline B}, \delta_0$, it is easy
to see that the dynamics is dominated by the slow modes, which enter with
coefficients $\exp(\pm i\varphi_s)$. The probabilities to find the pair in the states
$T_{+}$, $T_{-}$, and $S$ are given by
\begin{align}
\label{slowprobability}
|T_{+}(t)|^2=|T_{-}(t)|^2&=\frac{1}{2}\sin^2 \big[\varphi_s(t)-\varphi_s(t_0)\big],\nonumber \\
|S(t)|^2&=\cos^2 \big[\varphi_s(t)-\varphi_s(t_0)\big].
 \end{align}
To the leading order, the state $T_0$ is not involved into the dynamics. Similarly to Eq. (\ref{sprobability}), we can simplify the above expression using the smallness of $\nu$
as follows
\begin{equation}
|S(t)|^2=1-\nu^2\Bigg[\ln\left(\frac{1-\cos \omega t}{1-\cos \omega t_0}\right)-\ln\left(\frac{1+\cos \omega t}{1+\cos \omega t_0}\right)\Bigg]^2.
\end{equation}
Naturally, the magnitudes of the spectral harmonics are still given by Eq. (\ref{sfourier}).

\section{Decay of eigenmodes due to recombination}

\subsection{Qualitative consideration}

In calculating the decay, one should distinguish two limits: (i) very long recombination time, when recombination
does not affect the mode structure, so that all the components of the spinor, describing a given mode,  decay at the same rate, and (ii) short recombination time, when the mode structure
is completely modified by recombination. Qualitatively, it seems obvious that the slow modes
are affected by recombination stronger than the fast modes. This is because the fast modes
have a small singlet admixture.

In the absence of drive, the decay rate of the modes depends on the dimensionless product $\delta_0\tau$. This is physically apparent, since $S$ is coupled only to $T_0$, while $T_0$
is coupled only to $S$.
Then $\delta_0^{-1}$
is the time of $S-T_0$ beating, so that $\delta_0\tau$ is the number of oscillations before
the decay takes place. In the presence of a strong drive, $S$ is still coupled to $T_0$ only, see
Eq. (\ref{drivensystem3}), but, rather than  returning back to $S$, the $T_0$ component gives rise
to $T_{+}+T_{-}$, see Eq. (\ref{drivensystem4}). Moreover, it follows from Eq. (\ref{differential1})
that $S$ is effectively coupled to $\left(T_{+}-T_{-}\right)$, and the coupling coefficient  oscillates strongly with time. Correspondingly, $\left(T_{+}-T_{-}\right)$ is coupled back to $S$ by the oscillating coupling coefficient. The effect of such a nontrivial beating on
the decay of $\left(T_{+}-T_{-}\right)$ can be understood only from the quantitative analysis, which is presented below.

\subsection{Decay of slow modes}
To incorporate finite recombination time, we search for the solution of the system
Eqs. (\ref{differential1}), (\ref{differential2}) in the  form
\begin{eqnarray}
\label{SlowSemiclassics}
\begin{pmatrix}
  S \\ T_{+}-T_{-} \end{pmatrix}= \begin{pmatrix} \alpha \\ \beta
\end{pmatrix}\exp\left(\int\limits_{0}^{t} dt' \chi(t')\right).
\end{eqnarray}
Substituting this form into the system, we obtain

\begin{align}
\label{SlowSemiclassics1}
\alpha\Bigl(\chi^2+{\dot\chi}+\frac{\chi}{\tau}+\delta_0^2\Bigr)&=-2i\beta\chi B_1\frac{\delta_0}{\overline{B}}
\sin\omega t,\\
\beta\Big(\chi^2+{\dot\chi}+{\overline B}^2)&=-2i\alpha\Bigl(\chi+\frac{1}{\tau}\Bigr)B_1\frac{\overline{B}}{\delta_0}
\sin\omega t.
\end{align}
Multiplying the two equations yields

\begin{equation}
\label{SlowSemiclassics2}
\chi\left(\chi+\frac{1}{\tau}\right)=-\frac{\left(\delta_0^2+{\dot\chi}\right)\left(\chi^2+{\dot\chi}+{\overline B}^2\right)}{\overline{B}^2+4B_1^2\sin^2\omega t+\chi^2+{\dot\chi}}.
\end{equation}
The real part of $\chi(t)$ is responsible for the recombination-induced decay. In general, $\text{Re}~\hspace{-1mm}\chi(t)$ contains a constant part
and the part oscillating with a period $\pi/\omega$. Then it is convenient to express the effective decay time as
\begin{equation}
\label{taueffective}
\frac{1}{\tau_{\s eff}}=-\frac{\omega}{\pi}\int\limits_{0}^{\pi/\omega}\hspace{-1mm} dt~ \text{Re}~\hspace{-1mm}\chi(t).
\end{equation}
The right-hand side in Eq. (\ref{SlowSemiclassics2}) contains $B_1^2$ in denominator. Since $B_1$ is the biggest scale in the problem, the
left-hand side should be small. This, in-turn, suggests that either $\chi$ is small or $\chi$ is close to $-1/\tau$.
Small-$\chi$ solution
corresponds to the decay of the $T$-mode, while $\chi\approx -1/\tau$ solution  corresponds to the decay of the $S$-mode. Below we study the
two cases separately.

\subsubsection{Decay of S-mode}

For solution close to $\chi=-1/\tau$ we set
\begin{equation}
\chi_1=\chi+\frac{1}{\tau},
\end{equation}
where $\chi_1\ll 1/\tau$ is a correction. The function $\chi_1(t)$ is
responsible for the correction
\begin{equation}
\label{deltataueff}
\frac{\delta\tau_{eff}}{\tau^2}=-\frac{\omega}{\pi}\int\limits_{0}^{\pi/\omega}\hspace{-1mm} dt~\chi_1(t)
\end{equation}
to lifetime of the $S$-mode due to the coupling to the $T$-mode.

Smallness of $\chi_1$ allows
to replace $\chi(\chi+1/\tau)$ in the left-hand side by
$-\chi/\tau$. We can also replace the combination $\chi^2+{\dot\chi}$
in the right-hand side by $1/\tau^2$. The justification of this step
will be given later. After these simplifications Eq. (\ref{SlowSemiclassics2}) reduces to the following linear differential
equation for $\chi_1$
\begin{equation}
\label{SlowSemiclassicsnew2}
\chi_1=\frac{\tau\left(\delta_0^2+{\dot\chi_1}\right)\left({\overline B}^2+\frac{1}{\tau^2}\right)}{\overline{B}^2+\frac{1}{\tau^2}+4B_1^2\sin^2\omega t}.
\end{equation}
The solution of this equation reads
\begin{equation}
\label{SlowSemiclassicsolution}
\chi_1(t)=\delta_0^2\exp\Big[{-{\cal F}(t)}\Big]\int\limits_{-\infty}^{t}dt'\exp\Big[{{\cal F}(t')}\Big],
\end{equation}
where the function ${\cal F}(t)$ is defined as
\begin{equation}
\label{functiondefinition}
{\cal F}(t)=\frac{1}{\tau}\int\limits_{0}^{t}dt_1 \left(1+\frac{4B_1^2}{\overline{B}^2+\frac{1}{\tau^2}}\sin^2 \omega t_1\right).
\end{equation}
The form of the function ${\cal F}(t)$
depends on the dimensionless parameter $\lambda$ defined as
\begin{equation}
\label{ratio}
\lambda=\frac{4B_1^2}{\left(\overline{B}^2+\frac{1}{\tau^2}\right)\omega \tau}.
\end{equation}
For $\lambda \ll 1$, as we will see below, all times from $t=0$ to $t=\pi/\omega$ contribute to the decay Eq. (\ref{taueffective}). Then
one can replace $\sin^2\omega t$ by $1/2$ and obtain
\begin{equation}
\label{BigF1}
{\cal F}(t)\Big|_{\lambda \ll 1}\approx\frac{t}{\tau}\left(1+\frac{2B_1^2}{\overline{B}^2+\frac{1}{\tau^2}}\right).
\end{equation}
With ${\cal F}(t)$ being a simple exponent the integration in
Eq. (\ref{SlowSemiclassicsolution}) can be easily performed.
One concludes that $\chi_1(t)$ is a constant, so that the integration
in Eq. (\ref{deltataueff}) simply reduces to multiplication by $\pi/\omega$, and one gets
\begin{equation}
\label{resultcorrection1}
\frac{\delta\tau_{eff}}{\tau^2}= \frac{\delta_0^2\tau}{1+\frac{2B_1^2}{\overline{B}^2+\frac{1}{\tau^2}}}\approx \frac{2\delta_0^2}{\lambda \omega}.
\end{equation}
In the opposite limit $\lambda \gg 1$ only the times $t \ll \pi/\omega$ contribute to the decay. This allows to expand $\sin\omega t$ and yields the following form  of  ${\cal F}(t)$
\begin{equation}
\label{BigF2}
{\cal F}(t)\Big|_{\lambda \gg 1}\approx\frac{t}{\tau}+\frac{2B_1^2\omega^2t^3}{3\left(\overline{B}^2+\frac{1}{\tau^2}\right)\tau}.
\end{equation}
Substituting this form into Eq. (\ref{SlowSemiclassicsolution}) we get
\begin{widetext}
\begin{equation}
\label{wideequation}
\chi_1(t)=\delta_0^2\exp\Bigg[ -\frac{t}{\tau}-\frac{2B_1^2\omega^2t^3}{3\left(\overline{B}^2+\frac{1}{\tau^2}\right)\tau}\Bigg]
\Bigg\{\int\limits_{-\infty}^{t}dt'\exp\Bigg[ \frac{t'}{\tau}+\frac{2B_1^2\omega^2t'^3}{3\left(\overline{B}^2+\frac{1}{\tau^2}\right)\tau}\Bigg]\Bigg\}.
\end{equation}
\end{widetext}
The second term in the exponent can be presented as $\frac{2}{3}\lambda\omega^3t^3$. Thus, the characteristic time of the
change of this term is $1/\omega\lambda^{1/3}$, which is much smaller
than $1/\omega$, as we assumed above. Moreover, the conditions $\lambda \gg 1$ and $\omega\tau \gg 1$, guarantee that this time is much shorter than $\tau$. This allows to neglect $t/\tau$ in the exponent.
From this we conclude that the major contribution to the integral in Eq. (\ref{deltataueff}) comes from small times $\sim 1/\omega\lambda^{1/3}$, so that the upper limit can be replaced by $\infty$. Upon switching to dimensionless variable $\lambda^{1/3}\omega t$, we arrive to the final result
\begin{equation}
\label{resultcorrection2}
\frac{\delta\tau_{eff}}{\tau^2}=\frac{2^{2/3}\delta_0^2}{\pi\lambda^{2/3}\omega}I,
\end{equation}
where $I$ is a number defined as

\begin{equation}
\label{calI}
I=\int\limits_{0}^{\infty}dx ~ e^{-x^3/3}\int\limits_{-\infty}^{x}dy ~e^{y^3/3}=\frac{2}{3^{4/3}} {\mathrm \Gamma} \left(\frac{1}{3}\right)^2\approx 3.31.
\end{equation}

The results Eq. (\ref{resultcorrection1}) and Eq. (\ref{resultcorrection2}) differ by a factor $\lambda^{1/3}$, which means that, as the drive increases, we cross over from faster to slower decrease of $\chi_1$ with $\lambda$.  The physical reason for this is that for stronger drive the decay takes place during narrower time intervals around moments when $\sin \omega t=0$.

\subsubsection{Decay of T-mode}

The results (\ref{resultcorrection1}), (\ref{resultcorrection2}) are small corrections to the decay rate, $1/\tau$, of the $S$-mode. To get the decay rate of $T$-mode we set $\chi \ll 1/\tau$ in the differential equation (\ref{SlowSemiclassicsnew2}), which assumes the form
\begin{equation}
\label{SlowerSemiclassics}
\chi_{\s T}=-\frac{\tau\left(\delta_0^2+{\dot\chi_{\s T}}\right){\overline B}^2}{\overline{B}^2+4B_1^2\sin^2\omega t}.
\end{equation}
 We realize that this equation has the same form as Eq. (\ref{SlowSemiclassicsnew2}), only $\overline{B}^2+\frac{1}{\tau^2}$ is replaced by $\overline{B}^2$. Thus, the calculation of the decay of $T$-mode  is fully analogous to the above calculation. The result reads
\begin{multline}
\frac{1}{\tau_{eff}^{\s (T)}}=\frac{\delta_0^2\tau}{2}\left(\frac{\overline{B}}{B_1}\right)^2,~B_1 \ll \overline{B}(\omega \tau)^{1/2},\label{resultcorrection3}
\end{multline}
\begin{multline}
\label{resultcorrection4}
\frac{1}{\tau_{eff}^{\s (T)}}= \frac{\delta_0^2\tau}{2^{2/3}\pi\left(\omega \tau\right)^{1/3}}\left(\frac{\overline{B}}{B_1}\right)^{4/3},~ B_1 \gg \overline{B}(\omega \tau)^{1/2}.
\end{multline}
Note that Eqs. (\ref{resultcorrection3}), (\ref{resultcorrection4}) apply when $\delta_0 \tau \ll 1$. Physically, this condition means that the time of recombination is much smaller than time of ``talking" between $S$ and $T$, which is $\sim \delta_0^{-1}$.
\begin{figure}
\centering
\includegraphics[scale=0.30]{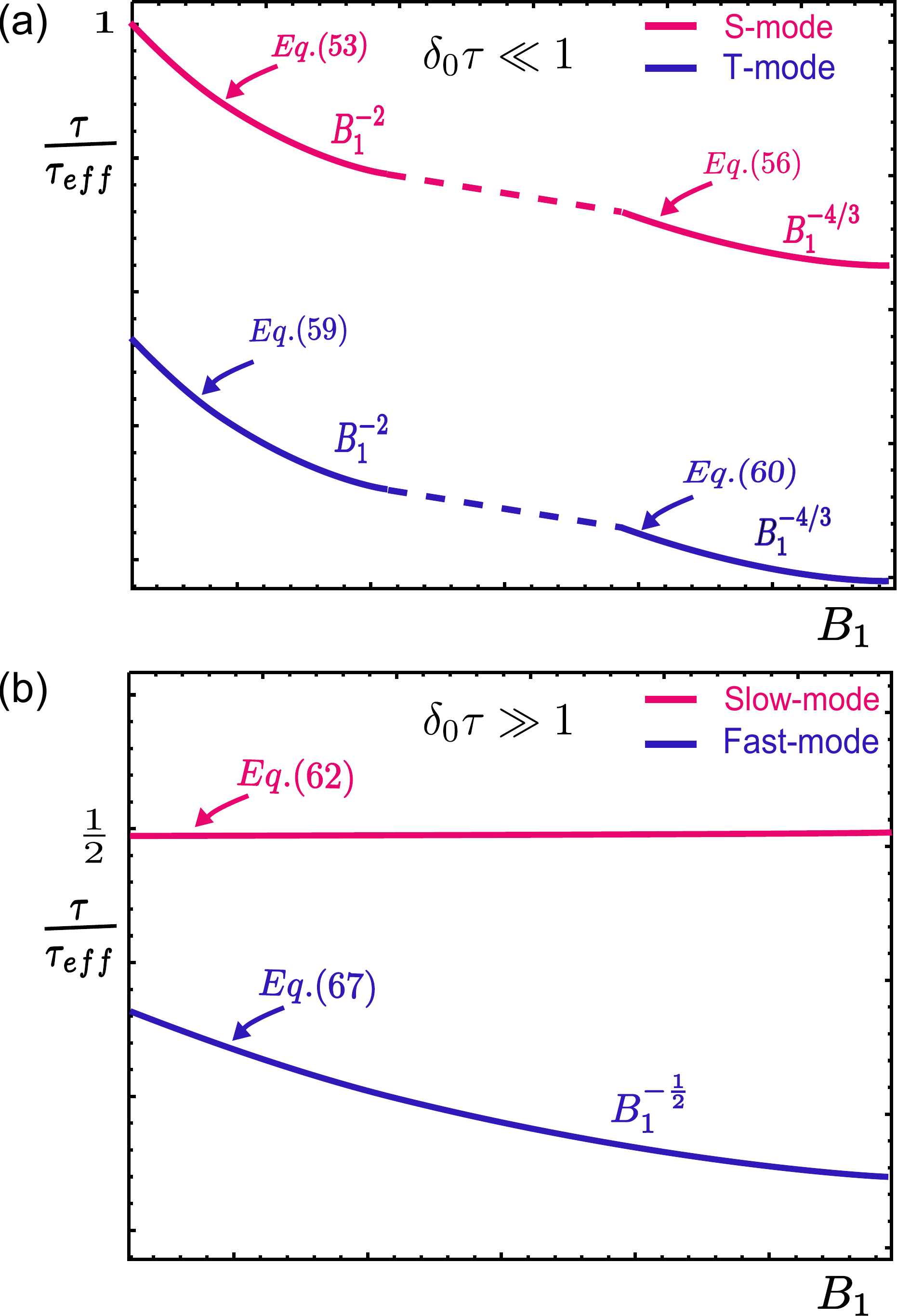}
\caption{(Color online) (a) Effective lifetimes $\tau_{\s eff}$ of the slow modes of $S$-type (red) and $T$-type (blue) are shown schematically versus the drive amplitude. The plots correspond to short recombination times, $\delta_0\tau \ll 1$, where $\delta_0$ is the difference of hyperfine fields acting on the pair-partners. At smaller values of $B_1$ recombination during entire periods of drive contributes to $\tau_{\s eff}$, while at larger $B_1$ the contribution
comes only from domains where the drive field passes through zero.
(b) For long recombination times, $\delta_0\tau \ll 1$, two slow modes are the combinations $S$ and $T$ and decay with same rate $\tau_{\s eff}^{-1}=1/2\tau$.   The decay rate of the fast modes is determined by Eq. (\ref{taueffectivefast1}). This decay takes place only during narrow intervals near $\omega t =\pi n$ when fast modes acquire an $S$-component.}
\label{p5}
\end{figure}
 In the opposite limit, $\delta_0 \tau \gg 1$, the modes of $S$-type and $T$-type decay collectively with decay rate $1/2\tau$, see Fig. \ref{p5}b. This can be seen from Eq. (\ref{SlowSemiclassics2}), which in this limit simplifies to
\begin{equation}
\label{collectiveequation}
\chi\left(\chi+\frac{1}{\tau}\right)=-\frac{\delta_0^2{\overline B}^2}{\overline{B}^2+4B_1^2\sin^2\omega t}.
\end{equation}
Two solutions of this equation have the form
\begin{equation}
\label{collectivesolution}
\chi=-\frac{1}{2\tau}\pm i\frac{\partial \varphi_{s}}{\partial t},
\end{equation}
where $\varphi_{s}$ is the phase of the slow mode defined by Eq. (\ref{slowphase}).

\subsection{Decay of fast modes}

Fast modes manifest themselves in Eq. (\ref{SlowSemiclassics2})
via a zero in the denominator in the right-hand side. Indeed, as
follows from Eq. (\ref{fastphase}), to the leading order,
$\partial\varphi_{\s f}/\partial t$ is equal to
$2iB_1\sin \omega t$. Thus, upon setting $\chi=\pm 2iB_1\sin \omega t$, two leading terms in denominator cancel each other.
Our goal is to find a real part of the correction to $\chi$,
caused by finite $\tau$. This correction is proportional to $1/\tau$. To accomplish this goal, it is convenient to rewrite Eq. (\ref{SlowSemiclassics2}) in the form

\begin{equation}
\label{fastequationdecay}
\overline{B}^2+4B_1^2\sin^2 \omega t +\chi^2 +\dot{\chi}=-\frac{\left(\delta_0^2+\dot{\chi}\right)\left(\chi^2+\dot{\chi}+\overline{B}^2\right)}{\chi\left(\chi+\frac{1}{\tau}\right)}.
\end{equation}
Expanding denominator in the right-hand side with respect to $1/\tau$, and taking all the $\tau$ independent terms to the left, we get
\begin{align}
\label{fastequationdecay1}
&\delta_0^2+\overline{B}^2+4B_1^2\sin^2 \omega t +\chi^2+ 2\dot{\chi}+ \frac{\left(\delta_0^2+\dot{\chi}\right)\left(\dot{\chi}+\overline{B}^2\right)}{\chi^2} \nonumber \\&=-\frac{\delta_0^2+\dot{\chi}}{\tau\chi}.
\end{align}
In the limit $\tau \rightarrow \infty$ the solution for $\chi$ fully reproduces the result Eq. (\ref{SlowSemiclassics}) including $\sin \omega t$ in the denominator. This can be verified by substituting
\begin{eqnarray}
\label{kifast}
\chi=\frac{\partial}{\partial t}\left(\varphi_{\s f}+\ln \sin \omega t\right)
\end{eqnarray}
into Eq. (\ref{fastequationdecay1}), where $\varphi_{\s f}$ is defined by Eq. (\ref{fastphase}).
Upon this substitution, the terms containing $B_1^2$ and $B_1$ get cancelled. Expanding the left-hand side around this solution yields the
sought correction to $\chi$. The real part of this correction has the form
\begin{equation}
\label{fastdecayrate}
\text{Re}~\chi_{\s f}=-\frac{1}{\tau}\Bigg[\frac{\delta_0^2}{4B_1^2\sin^2 \omega t}\Bigg],
\end{equation}
where we have neglected to corrections to $B_1^2 \sin^2 \omega t$ in the denominator. The reason is that semiclassical   condition Eq. (\ref{condition}) guarantees that these corrections are small.

Substituting $\text{Re}~\chi_{\s f}$ into the definition
Eq. (\ref{taueffective}) of the effective decay time,
we realize that the integral comes from small times,
so that the  cutoff, $t_{\text{min}}$, is set by the applicability of the semiclassics.
Thus, with the accuracy of a numerical factor one gets
\begin{align}
\label{taueffectivefast1}
\frac{1}{\tau_{\s eff}}\approx \frac{\delta_0^2}{4B_1^2\omega^2 t_{\text{min}}\tau}
=\frac{\delta_0^2}{\left(2B_1\overline{B}\right)^{1/2}\omega\tau}.
\end{align}
The falloff of the decay rate with $B_1$ is illustrated in Fig. \ref{p5}b.

\begin{figure}
\includegraphics[scale=0.40]{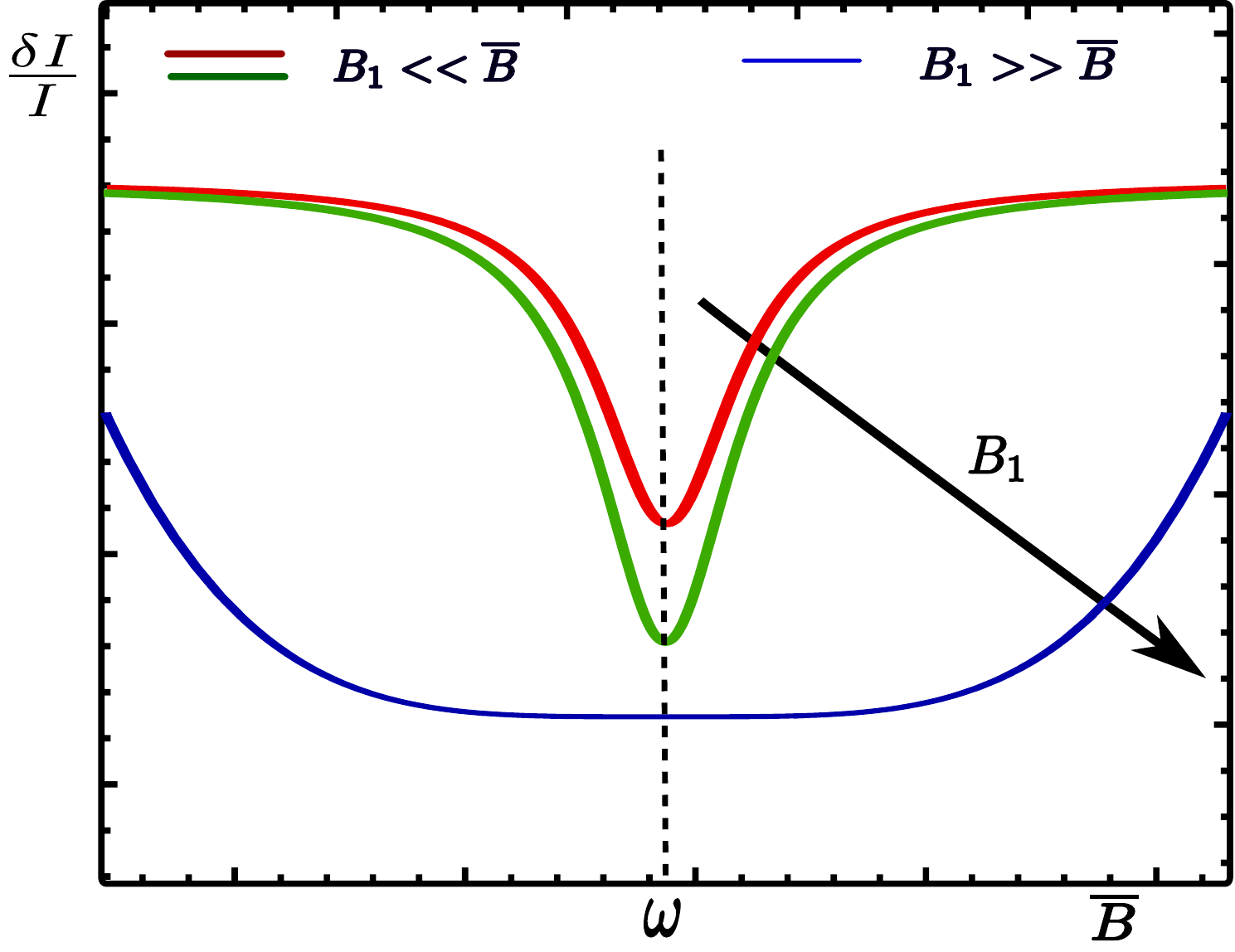}
\caption{(Color online)
The drive-induced relative correction to the current, $\delta I/I$, versus the dc magnetic field, ${\overline B}$, is illustrated schematically in different domains of the ac drive amplitude $B_1$. For weak drive, $B_1\ll {\overline B}$, the current is sensitive only to the resonant drive\cite{we3,Boehme1,Baker} with $\omega\approx
{\overline B}$. The widths of the curves, $|\omega -{\overline B}|\sim \delta_0$,  depends on $B_1$ only weakly, while the magnitude of $\delta I$ grows linearly with $B_1$ following the formation of long-living modes. For strong drive, the recombination from long-living modes is suppressed even stronger, while the sensitivity to the drive frequency becomes weak. }
\label{p7}
\end{figure}

\section{Discussion}
 Our prediction for the behavior of organic magnetoresistance in the regime of a strong
drive is the following. In the weak-drive regime the drive first enhances recombination,
but, upon increasing of $B_1$, recombination is slowed down on average. This is due to
formation of  three long-living modes\cite{we3,Boehme1,Bayliss}: $\frac{1}{\sqrt{2}}\left(T_{+}-T_{-}\right)$, and $\frac{1}{2}\left(T_{+}\pm \sqrt{2}T_0+  T_{-}\right)$.
The change of the average recombination rate is reflected in the magnitude of current.
Importantly, the sensitivity to the drive emerges only near the resonance condition ${\overline B}=\omega$. Our study shows that, under the strong drive, the above long-living modes become even more long-living, see Eqs. (\ref{resultcorrection3}), (\ref{resultcorrection4}), (\ref{taueffectivefast1}). At the same time, this suppression of
recombination takes place for arbitrary relation between ${\overline B}$ and $\omega$. Thus, a dip in the current vs. magnetic field dependence  at weak drive\cite{Baker,Boehme1} should transform into a {\em broad plateau} at strong drive. This is illustrated in Fig. \ref{p7}.

It is instructive to compare the spin dynamics of a pair
under a strong drive to the conventional  spin dynamics of a
pair under a weak resonant drive. In the latter case,\cite{Recombination4,Araki,Gorelik} the spinor
components are the combinations of $\cos B_1t$ and $\cos 2B_1t$.
Harmonics $\cos B_1t$ corresponds to one pair partner involved in the Rabi oscillations,
while the harmonics $\cos 2B_1t$ correspond to both partners involved into the Rabi oscillations.
Loosely speaking, the outcome of our study is that, with strong drive, the argument $B_1t$ should be replaced by $\frac{2B_1}{\omega}\cos \omega t$, and correspondingly, the
argument $2B_1t$ should be replaced by $\frac{4B_1}{\omega}\cos \omega t$. This gives
rise to the Fourier spectrum with numerous harmonics.
If the hyperfine fields for the pair partners are different, the dynamics of a weakly
driven pair contains a harmonics, corresponding to the difference of the Rabi frequencies.
For a strong drive, this harmonics evolves into the strongly anharmonic slow mode, see Eqs.
(\ref{slowphase}), (\ref{sfourier}).

One of the outcomes of our study is that, when the recombination
time from $S$ is short, then the decay time of $\frac{1}{\sqrt{2}}\left(T_{+}-T_{-}\right)$
mode is long.
Similar effect takes place even in the absence of drive. Indeed, setting
$B_1=0$ in Eq. (\ref{differential1}), we get $S\propto \exp(-\tilde{\chi}t)$, where
\begin{equation}
\label{tilde}
\tilde{\chi}=\frac{1}{2\tau}\pm \left(\frac{1}{4\tau^2}-\delta_0^2    \right)^{1/2}.
\end{equation}
We see that, when $\delta_0\tau \ll 1$, the two decay rates are strongly  different\cite{we,we1}, and are equal to $1/\tau$ and $\delta_0^2\tau$.
Thus, the faster is the recombination from $S$, the slower is the decay of the mode
$T_0$ to which $S$ is coupled. The decay rate $\delta_0^2\tau$ has the same form
as the decay of magnetic resonance in the limit of spectral narrowing or the Dyakonov-Perel
spin relaxation time.

In the presence of a weak resonant drive with ${\overline B}\approx \omega$ and $B_1\ll \omega$ short $\tau$ also leads to a long-living mode\cite{we3}, but this time it is
the mode $\frac{1}{\sqrt{2}}\left(T_{+}-T_{-}\right)$. Our finding in the present paper is that, for
strong drive, the decay of $\frac{1}{\sqrt{2}}\left(T_{+}-T_{-}\right)$ is suppressed even stronger.

%

\section{Concluding remarks}
(i). In calculating the Fourier spectrum of the fast modes we neglected the corrections to
spinors proportional to $1/\sin \omega t$ and to $\cos \omega t/ \sin^2 \omega t$.
The Fourier integral of the first correction diverges logarithmically as $\int dt e^{i n\omega t}/\sin\omega t$.
the Fourier integral of the second correction, after integration by parts,
reduces to $in\int dt e^{i n\omega t}/\sin\omega t$. The upper cutoff of the logarithm is
$t_{\text{max}}\sim 1/n\omega$, while the lower cutoff $t_{\text{min}}\sim 1/\left(B_1\omega   \right)^{1/2}$ is set by the condition Eq. (\ref{condition}).
The maximum number of the Fourier component for which  $t_{\text{max}}$ exceeds $t_{\text{min}}$ is $n\sim \left(B_1/\omega\right)^{1/2}$.
Since $\cos \omega t/ \sin^2 \omega t$ correction enters into the spinor Eq.
(\ref{fastmatrix}) with a small coefficient $\omega/B_1$, we conclude that neglecting this
correction was justified.

(ii). Throughout the paper we assumed that the spin dynamics takes place outside the
intervals $\omega t-\pi n \sim  \left(B_1/\omega\right)^{1/2}$. The system passes these
intervals by performing the Landau-Zener transitions. We can now express the transition
probability via the drive magnitude, as $1-\exp{(-2\pi {\overline B}^2/\omega B_1)}$.
Therefore, for strong drive, this probability is small.

(iii). In Landau-Zener-Stueckelberg interferometry\cite{LZ3} the experimentally
varied parameters are the flux amplitude and frequency, but also the
baseline of the magnetic flux. A lively response is observed as a function
of the ratio of the amplitude and the baseline flux.
 In our language, the baseline corresponds to a constant magnetic field along the $x$-axis. Thus, the similar interplay can be achieved by simply
 rotating the dc field.

\noindent{\em Acknowledgements.}
We are grateful to C. Boehme for piquing
our interest in the subject.
This work was supported by NSF through MRSEC DMR-1121252.

\end{document}